\pdfoutput=1
\documentclass[12pt,a4paper]{article}

\usepackage{ifthen} 
\usepackage{etoolbox} 
\usepackage{forarray} 
\newboolean{pdflatex}
\setboolean{pdflatex}{true} 

\newboolean{articletitles}
\setboolean{articletitles}{true} 

\newboolean{uprightparticles}
\setboolean{uprightparticles}{false} 

\def\paperauthors{LHCb collaboration} 
\def\paperasciititle{Observation of the Bs2DstDst decay} 
\def\papertitle{Observation of the \decay{\Bs}{\Dstarp\Dstarm} decay} 
\def\paperkeywords{{High Energy Physics}, {LHCb}} 
\def\papercopyright{\the\year\ CERN for the benefit of the LHCb collaboration} 
\def\paperlicence{CC BY 4.0 licence}
\def\paperlicenceurl{https://creativecommons.org/licenses/by/4.0/}

\usepackage[top=1in, bottom=1.25in, left=1in, right=1in]{geometry}

\usepackage{microtype}
\usepackage{lineno}  
\usepackage{xspace} 
\usepackage{caption} 

\usepackage{graphicx}  
\usepackage{color}
\usepackage{colortbl}
\graphicspath{{./figs/}} 

\usepackage{amsmath} 
\usepackage{amssymb}
\usepackage{amsfonts}
\usepackage{upgreek} 

\newcommand*\patchAmsMathEnvironmentForLineno[1]{%
\expandafter\let\csname old#1\expandafter\endcsname\csname #1\endcsname
\expandafter\let\csname oldend#1\expandafter\endcsname\csname
end#1\endcsname
 \renewenvironment{#1}%
   {\linenomath\csname old#1\endcsname}%
   {\csname oldend#1\endcsname\endlinenomath}%
}
\newcommand*\patchBothAmsMathEnvironmentsForLineno[1]{%
  \patchAmsMathEnvironmentForLineno{#1}%
  \patchAmsMathEnvironmentForLineno{#1*}%
}
\AtBeginDocument{%
\patchBothAmsMathEnvironmentsForLineno{equation}%
\patchBothAmsMathEnvironmentsForLineno{align}%
\patchBothAmsMathEnvironmentsForLineno{flalign}%
\patchBothAmsMathEnvironmentsForLineno{alignat}%
\patchBothAmsMathEnvironmentsForLineno{gather}%
\patchBothAmsMathEnvironmentsForLineno{multline}%
\patchBothAmsMathEnvironmentsForLineno{eqnarray}%
}

\usepackage{hyperxmp}

\usepackage[pdftex,
            pdfauthor={\paperauthors},
            pdftitle={\paperasciititle},
            pdfkeywords={\paperkeywords},
            pdfcopyright={Copyright (C) \papercopyright},
            pdflicenseurl={\paperlicenceurl}]{hyperref}

\usepackage{xargs}     
\usepackage[colorinlistoftodos,prependcaption,textsize=scriptsize]{todonotes}
\newcommandx{\unsure}[2][1=]{\todo[linecolor=red,backgroundcolor=red!25,bordercolor=red,#1]{#2}}
\newcommandx{\change}[2][1=]{\todo[linecolor=blue,backgroundcolor=blue!25,bordercolor=blue,#1]{#2}}
\newcommandx{\info}[2][1=]{\todo[linecolor=OliveGreen,backgroundcolor=OliveGreen!25,bordercolor=OliveGreen,#1]{#2}}
\newcommandx{\improvement}[2][1=]{\todo[linecolor=Plum,backgroundcolor=Plum!25,bordercolor=Plum,#1]{#2}}
\newcommandx{\thiswillnotshow}[2][1=]{\todo[disable,#1]{#2}}

\usepackage[bottom,flushmargin,hang,multiple]{footmisc}

\usepackage[all]{hypcap} 

\usepackage{xspace} 
\usepackage{upgreek}

\def\lhcb   {\mbox{LHCb}\xspace}

\def\MagUp {\mbox{\em Mag\kern -0.05em Up}\xspace}

\ifthenelse{\boolean{uprightparticles}}%
{

 \def\Ppi         {\ensuremath{\uppi}\xspace}

 \def\PDelta      {\ensuremath{\Delta}\xspace}                 
 \def\PXi         {\ensuremath{\Xi}\xspace}                 
 \def\PLambda     {\ensuremath{\Lambda}\xspace}                 
 \def\PSigma      {\ensuremath{\Sigma}\xspace}                 
 \def\POmega      {\ensuremath{\Omega}\xspace}                 
 \def\PUpsilon    {\ensuremath{\Upsilon}\xspace}
 \let\oldPi\Pi
 \def\PPi         {\ensuremath{\oldPi}\xspace}

 \def\PB      {\ensuremath{\mathrm{B}}\xspace}                 
                  
 \def\PD      {\ensuremath{\mathrm{D}}\xspace}

 \def\PK      {\ensuremath{\mathrm{K}}\xspace}

 \def\PW      {\ensuremath{\mathrm{W}}\xspace}

 \def\Pb      {\ensuremath{\mathrm{b}}\xspace}                 
 \def\Pc      {\ensuremath{\mathrm{c}}\xspace}                 
 \def\Pd      {\ensuremath{\mathrm{d}}\xspace}

 \def\Pi      {\ensuremath{\mathrm{i}}\xspace}

 \def\Ps      {\ensuremath{\mathrm{s}}\xspace}

 \def\thebaroffset{0.0em}
}
{

 \def\Ppi         {\ensuremath{\pi}\xspace}

 \mathchardef\PDelta="7101
 \mathchardef\PXi="7104
 \mathchardef\PLambda="7103
 \mathchardef\PSigma="7106
 \mathchardef\POmega="710A
 \mathchardef\PUpsilon="7107
 \mathchardef\PPi="7105
                  
 \def\PB      {\ensuremath{B}\xspace}                 
                  
 \def\PD      {\ensuremath{D}\xspace}

 \def\PK      {\ensuremath{K}\xspace}

 \def\PW      {\ensuremath{W}\xspace}

 \def\Pb      {\ensuremath{b}\xspace}                 
 \def\Pc      {\ensuremath{c}\xspace}                 
 \def\Pd      {\ensuremath{d}\xspace}

 \def\Pi      {\ensuremath{i}\xspace}

 \def\Ps      {\ensuremath{s}\xspace}

 \def\thebaroffset{0.18em}
}
\newcommand{\offsetoverline}[2][\thebaroffset]{\kern #1\overline{\kern -#1 #2}}%

\makeatletter
\ifcase \@ptsize \relax
  \newcommand{\miniscule}{\@setfontsize\miniscule{4}{5}}
\or
  \newcommand{\miniscule}{\@setfontsize\miniscule{5}{6}}
\or
  \newcommand{\miniscule}{\@setfontsize\miniscule{5}{6}}
\fi
\makeatother

\DeclareRobustCommand{\optbar}[1]{\shortstack{{\miniscule (\rule[.5ex]{1.25em}{.18mm})}
  \\ [-.7ex] $#1$}}

\def\W      {{\ensuremath{\PW}}\xspace}

\def\dquark    {{\ensuremath{\Pd}}\xspace}

\def\squark    {{\ensuremath{\Ps}}\xspace}

\def\cquark    {{\ensuremath{\Pc}}\xspace}

\def\bquark    {{\ensuremath{\Pb}}\xspace}
\def\bquarkbar {{\ensuremath{\overline \bquark}}\xspace}
\def\bbbar     {{\ensuremath{\bquark\bquarkbar}}\xspace}

\def\pion   {{\ensuremath{\Ppi}}\xspace}

\def\pip    {{\ensuremath{\pion^+}}\xspace}
\def\pim    {{\ensuremath{\pion^-}}\xspace}

\def\kaon    {{\ensuremath{\PK}}\xspace}

\def\KorKbar {\kern \thebaroffset\optbar{\kern -\thebaroffset \PK}{}\xspace}

\def\Kp      {{\ensuremath{\kaon^+}}\xspace}
\def\Km      {{\ensuremath{\kaon^-}}\xspace}

\def\Dbar    {{\ensuremath{\offsetoverline{\PD}}}\xspace}
\def\D       {{\ensuremath{\PD}}\xspace}

\def\DorDbar {\kern \thebaroffset\optbar{\kern -\thebaroffset \PD}\xspace}
\def\Dz      {{\ensuremath{\D^0}}\xspace}
\def\Dzb     {{\ensuremath{\Dbar{}^0}}\xspace}
\def\Dp      {{\ensuremath{\D^+}}\xspace}
\def\Dm      {{\ensuremath{\D^-}}\xspace}

\def\DpDm    {\ensuremath{\Dp {\kern -0.16em \Dm}}\xspace}
\def\Dstar   {{\ensuremath{\D^*}}\xspace}

\def\Dstarp  {{\ensuremath{\D^{*+}}}\xspace}
\def\Dstarm  {{\ensuremath{\D^{*-}}}\xspace}

\def\B       {{\ensuremath{\PB}}\xspace}

\def\BorBbar {\kern \thebaroffset\optbar{\kern -\thebaroffset \PB}\xspace}

\def\Bd      {{\ensuremath{\B^0}}\xspace}

\def\BdorBdbar {\kern \thebaroffset\optbar{\kern -\thebaroffset \Bd}\xspace}

\def\Bs      {{\ensuremath{\B^0_\squark}}\xspace}

\def\BsorBsbar {\kern \thebaroffset\optbar{\kern -\thebaroffset \Bs}\xspace}

\def\Y#1S{\ensuremath{\PUpsilon{(#1S)}}\xspace}

\def\LorLbar     {\kern \thebaroffset\optbar{\kern -\thebaroffset \PLambda}\xspace}

\newcommand{\decay}[2]{\ensuremath{#1\!\to #2}\xspace} 

\def\to                 {\ensuremath{\rightarrow}\xspace}

\def\CP                {{\ensuremath{C\!P}}\xspace}

\newcommand{\betas}{{\ensuremath{\beta_{\squark}}}\xspace}

\def\AT#1     {\ensuremath{A_{\mathrm{T}}^{#1}}\xspace}           

\def\C#1      {\ensuremath{\mathcal{C}_{#1}}\xspace}                       
\def\Cp#1     {\ensuremath{\mathcal{C}_{#1}^{'}}\xspace}                    
\def\Ceff#1   {\ensuremath{\mathcal{C}_{#1}^{\mathrm{(eff)}}}\xspace}        
\def\Cpeff#1  {\ensuremath{\mathcal{C}_{#1}^{'\mathrm{(eff)}}}\xspace}       
\def\Ope#1    {\ensuremath{\mathcal{O}_{#1}}\xspace}                       
\def\Opep#1   {\ensuremath{\mathcal{O}_{#1}^{'}}\xspace}                    

\newcommand{\nospaceunit}[1]{\ensuremath{\text{#1}}}       
\newcommand{\aunit}[1]{\ensuremath{\text{\,#1}}}       

\newcommand{\tev}{\aunit{Te\kern -0.1em V}\xspace}
\newcommand{\gev}{\aunit{Ge\kern -0.1em V}\xspace}
\newcommand{\mev}{\aunit{Me\kern -0.1em V}\xspace}
\newcommand{\kev}{\aunit{ke\kern -0.1em V}\xspace}
\newcommand{\ev}{\aunit{e\kern -0.1em V}\xspace}
 
\newcommand{\mevc}{\ensuremath{\aunit{Me\kern -0.1em V\!/}c}\xspace}
\newcommand{\gevc}{\ensuremath{\aunit{Ge\kern -0.1em V\!/}c}\xspace}
\newcommand{\mevcc}{\ensuremath{\aunit{Me\kern -0.1em V\!/}c^2}\xspace}
\newcommand{\gevcc}{\ensuremath{\aunit{Ge\kern -0.1em V\!/}c^2}\xspace}

\def\mm   {\aunit{mm}\xspace}

\def\mum  {\ensuremath{\,\upmu\nospaceunit{m}}\xspace}

\def\fb   {\ensuremath{\aunit{fb}}\xspace}
\def\invfb   {\ensuremath{\fb^{-1}}\xspace}

\def\ps   {\ensuremath{\aunit{ps}}\xspace}

\newcommand{\chisq}{\ensuremath{\chi^2}\xspace}

\newcommand{\chisqip}{\ensuremath{\chi^2_{\text{IP}}}\xspace}

\def\gsim{{~\raise.15em\hbox{$>$}\kern-.85em
          \lower.35em\hbox{$\sim$}~}\xspace}
\def\lsim{{~\raise.15em\hbox{$<$}\kern-.85em
          \lower.35em\hbox{$\sim$}~}\xspace}

\def\PDF {PDF\xspace}

\def\pt         {\ensuremath{p_{\mathrm{T}}}\xspace}

\def\ptot       {\ensuremath{p}\xspace}

\def\evtgen     {\mbox{\textsc{EvtGen}}\xspace}

\def\geant      {\mbox{\textsc{Geant4}}\xspace}

\def\photos     {\mbox{\textsc{Photos}}\xspace}

\def\pythia     {\mbox{\textsc{Pythia}}\xspace}

\def\tell1  {TELL1\xspace}
\def\ukl1   {UKL1\xspace}

\newcommand{\eg}{\mbox{\itshape e.g.}\xspace}
\newcommand{\ie}{\mbox{\itshape i.e.}\xspace}

\newcommand{\lhcborcid}[1]{\href{https://orcid.org/#1}{\hspace*{0.1em}\raisebox{-0.45ex}{\includegraphics[width=1em]{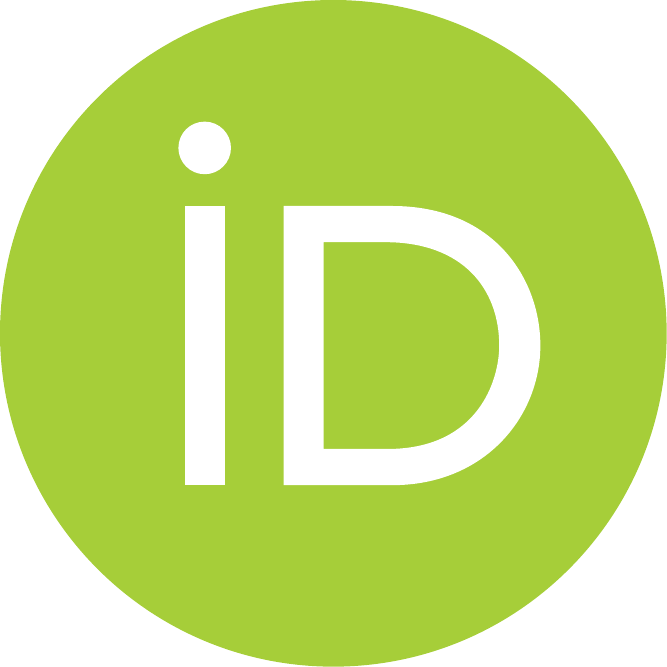}}}}

\usepackage{cite} 
\usepackage{mciteplus}

\def\BdToDstDst         {\ensuremath{\decay{\Bd}{\Dstarp\Dstarm}}\xspace}
\def\BsToDstDst         {\ensuremath{\decay{\Bs}{\Dstarp\Dstarm}}\xspace}

\def\BToDD         {\ensuremath{\decay{\B}{\D\D}}\xspace}
\usepackage{longtable} 
\usepackage[withpage,nolist]{acronym}

\usepackage{relsize}

\usepackage{cleveref}
\crefname{chapter}{Ch.\@}{Chs.\@}
\crefname{section}{Sec.\@}{Secs.\@}
\crefname{subsection}{Sec.\@}{Secs.\@}
\crefname{appendix}{Appendix\@}{Appendices\@}
\crefname{figure}{Fig.\@}{Figs.\@}
\crefname{table}{Table\@}{Tables\@}
\crefname{equation}{Eq.\@}{Eqs.\@}

\usepackage{nicefrac}

\usepackage{booktabs}
\usepackage{longtable}

\def\NumberWithoutMisIDRunOne   {\ensuremath{0.2}}
\def\NumberWithMisIDRunOne   {\ensuremath{1.4}}

\def\NumberWithoutMisIDRunTwo   {\ensuremath{1.3}}
\def\NumberWithMisIDRunTwo   {\ensuremath{2.5}}
\def\BsNominalRunOne   {\ensuremath{20\pm5}}
\def\BdNominalRunOne   {\ensuremath{251\pm16}}

\def\BsNominalRunTwo   {\ensuremath{79\pm10}}
\def\BdNominalRunTwo   {\ensuremath{1123\pm34}}

\def\EfficiencyRatioRunOne   {\ensuremath{1.037\pm0.016}}
\def\EfficiencyRatioRunTwo   {\ensuremath{1.082\pm0.012}}
\def\fsoverfdRunOne   {\ensuremath{0.239\pm0.008}}
\def\CandidatesRatioRunOne   {\ensuremath{0.081\pm0.020}}
\def\fsoverfdRunTwo   {\ensuremath{0.254\pm0.008}}
\def\CandidatesRatioRunTwo   {\ensuremath{0.071\pm0.010}}
\def\finalbrRunOneValue   {\ensuremath{0.328}}
\def\finalbrRunOneStat   {\ensuremath{0.080}}

\def\finalbrRunOneSystNoFsFd   {\ensuremath{0.016}}
\def\finalbrRunOneSystOnlyFsFd   {\ensuremath{0.010}}
\def\finalbrRunTwoValue   {\ensuremath{0.257}}
\def\finalbrRunTwoStat   {\ensuremath{0.035}}

\def\finalbrRunTwoSystNoFsFd   {\ensuremath{0.010}}
\def\finalbrRunTwoSystOnlyFsFd   {\ensuremath{0.008}}
\def\finalbrCombinedValue   {\ensuremath{0.269}}
\def\finalbrCombinedStat   {\ensuremath{0.032}}

\def\finalbrCombinedSystNoFsFd   {\ensuremath{0.011}}
\def\finalbrCombinedSystOnlyFsFd   {\ensuremath{0.008}}
\def\integratedcL   {\ensuremath{0.947}}
\def\integratedcH   {\ensuremath{1.045}}

\newcommand{\asymbol}[3]{
\ensuremath{#1
^{\ForEach{;}{
\ifboolexpr{test {\ifnumcomp{\the\thislevelcount}{=}{1}}}
  {\text{\thislevelitem}}
  {, \text{\thislevelitem}}
}{#2}}
_{\ForEach{;}{
\ifboolexpr{test {\ifnumcomp{\the\thislevelcount}{=}{1}}}
  {\text{\thislevelitem}}
  {, \text{\thislevelitem}}
}{#3}}
\xspace}
}

\newcommand{\pdf}[2]{\asymbol{\mathcal{P}}{#1}{#2}}

\newcommand{\yield}[2]{\asymbol{N}{#1}{#2}}

\begin{document}

\begin{acronym}[misID]
\acro{CKM}{Cabibbo–Kobayashi–Maskawa}
\acro{LHCb}[\lhcb]{Large Hadron Collider beauty}
\acro{PID}[PID]{particle identification}
\acro{IP}[IP]{impact parameter}
\acro{PV}[PV]{primary $pp$ collision vertex}
\acroplural{PV}[PVs]{primary vertices}
\acro{PDF}[\PDF]{probability density function}
\acroplural{PDFs}[\PDF]{probability density functions}
\end{acronym}

\acused{LHCb}
\renewcommand{\thefootnote}{\fnsymbol{footnote}}
\setcounter{footnote}{1}


\begin{titlepage}
\pagenumbering{roman}

\vspace*{-1.5cm}
\centerline{\large EUROPEAN ORGANIZATION FOR NUCLEAR RESEARCH (CERN)}
\vspace*{1.5cm}
\noindent
\begin{tabular*}{\linewidth}{lc@{\extracolsep{\fill}}r@{\extracolsep{0pt}}}
\ifthenelse{\boolean{pdflatex}}
{\vspace*{-1.5cm}\mbox{\!\!\!\includegraphics[width=.14\textwidth]{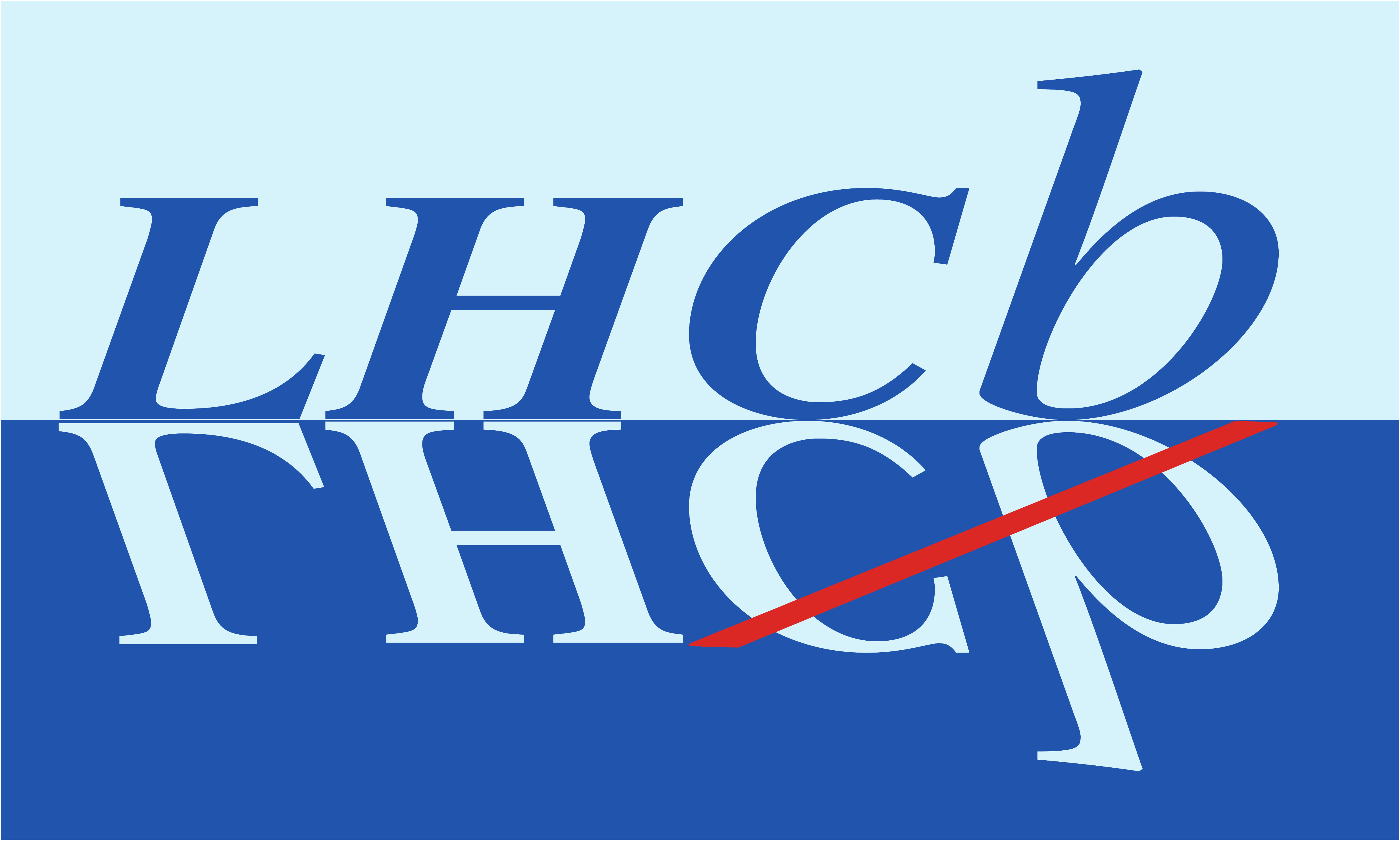}} & &}%
{\vspace*{-1.2cm}\mbox{\!\!\!\includegraphics[width=.12\textwidth]{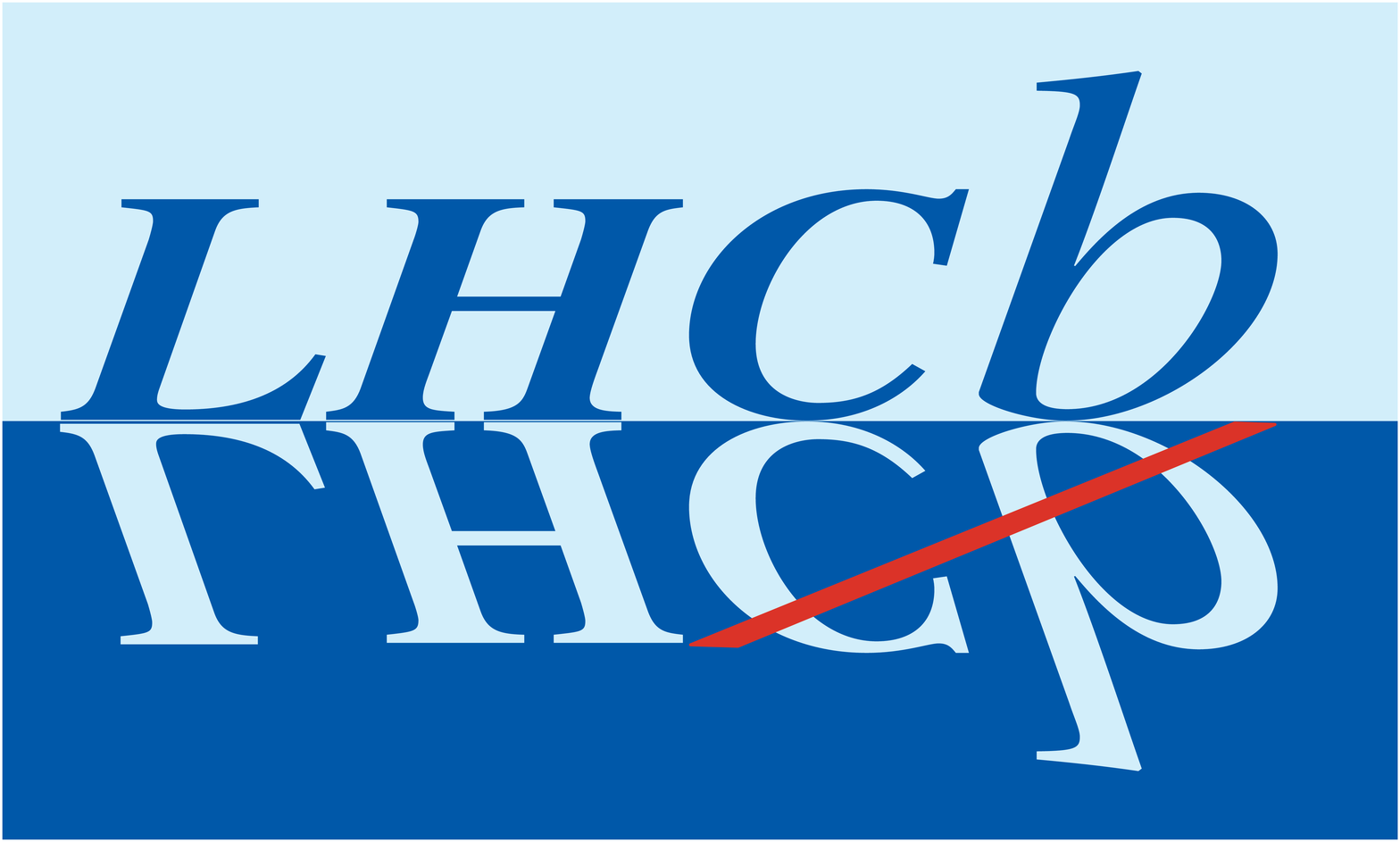}} & &}%
\\
 & & CERN-EP-2022-193 \\  
 & & LHCb-PAPER-2022-023 \\  
 & & July 13, 2023 \\ 
 & & \\
\end{tabular*}

\vspace*{4.0cm}

{\normalfont\bfseries\boldmath\huge
\begin{center}
  \papertitle 
\end{center}
}

\vspace*{1cm}

\begin{center}
\paperauthors\footnote{Authors are listed at the end of this paper.}
\end{center}

\vspace{\fill}

\begin{abstract}
\noindent
The first observation of the \BsToDstDst decay and the measurement of its branching ratio relative to the \BdToDstDst decay are presented.
The data sample used corresponds to an integrated luminosity of 9\,$\text{fb}^{-1}$ of proton-proton collisions recorded by the LHCb experiment at centre-of-mass energies of 7, 8 and 13 \,TeV between 2011 and 2018.
The decay is observed with more than $10$ standard deviations and the time-integrated ratio of branching fractions is determined to be
\begin{align*}
  \frac{\mathcal{B}(\decay{\Bs}{\Dstarp\Dstarm})}{\mathcal{B}(\decay{\Bd}{\Dstarp\Dstarm})} = \finalbrCombinedValue \pm \finalbrCombinedStat \pm \finalbrCombinedSystNoFsFd \pm \finalbrCombinedSystOnlyFsFd\, ,
\end{align*}
where the first uncertainty is statistical, the second systematic and the third due to the uncertainty of the fragmentation fraction ratio $f_\squark/f_\dquark$. 
The \BsToDstDst branching fraction is calculated to be
\begin{align*}
  \mathcal{B}(\decay{\Bs}{\Dstarp\Dstarm}) = (2.15 \pm 0.26 \pm 0.09 \pm 0.06 \pm 0.16)\times 10^{-4} \,,
\end{align*}
where the fourth uncertainty is due to the \BdToDstDst branching fraction. These results are calculated using the average \Bs meson lifetime in simulation. Correction factors are reported for scenarios where either a purely heavy or a purely light \Bs eigenstate is considered.
\end{abstract}

\vspace*{1cm}

\begin{center}
  Published in JHEP 07 (2023) 119
\end{center}

\vspace{\fill}

{\footnotesize 
\centerline{\copyright~\papercopyright. \href{\paperlicenceurl}{\paperlicence}.}}
\vspace*{2mm}

\end{titlepage}


\newpage
\setcounter{page}{2}
\mbox{~}
\newpage
%
%
%
%


\renewcommand{\thefootnote}{\arabic{footnote}}
\setcounter{footnote}{0}



\pagestyle{plain} 
\setcounter{page}{1}
\pagenumbering{arabic}


\section{Introduction}
\label{sec:Introduction}
\begin{figure}[hbp]
    \centering
    \includegraphics[width=\textwidth]{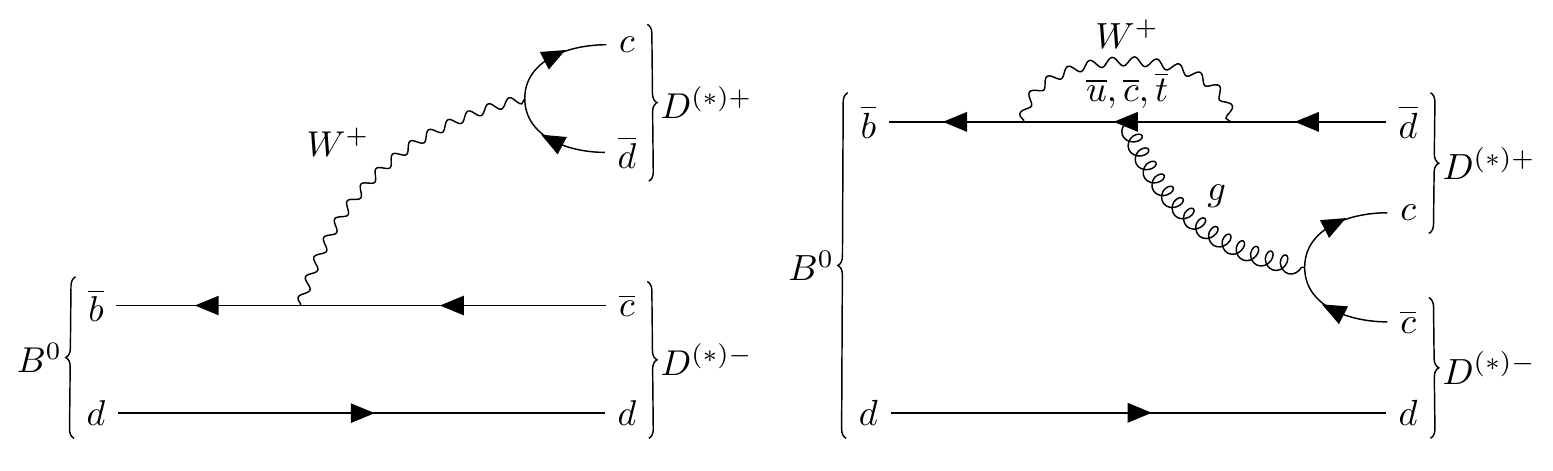}\vspace{1em}
    \includegraphics[width=\textwidth]{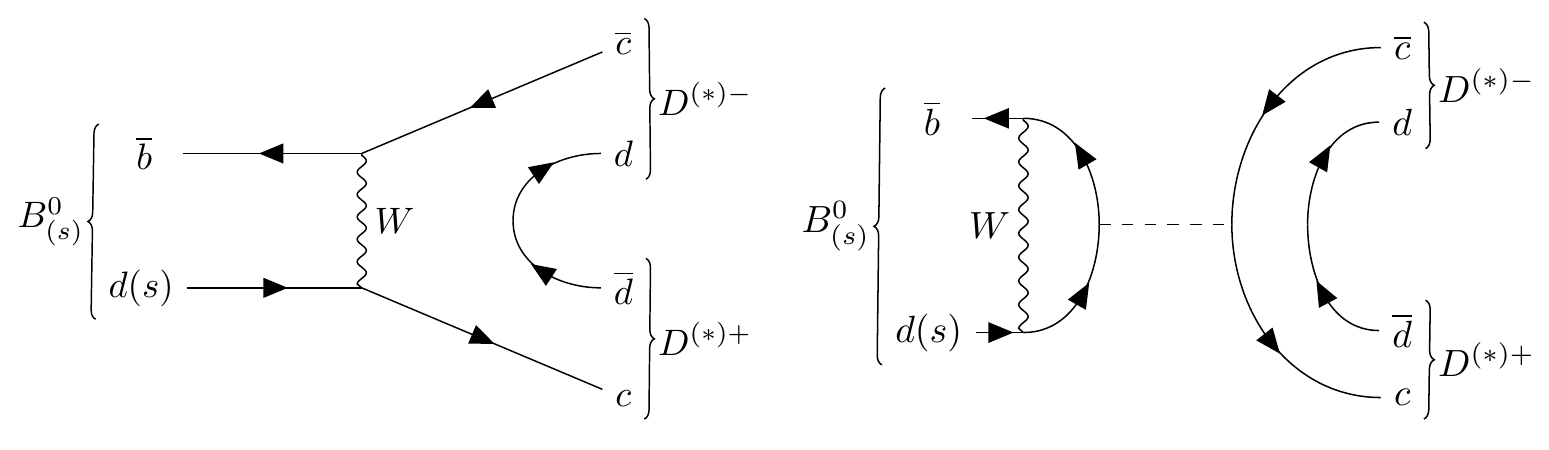}
    \caption{Dominant Feynman diagrams for \BToDD decays comprising (top left) tree, (top right) penguin, (bottom left) \W-exchange and (bottom right) penguin-annihilation transitions.
    The \BsToDstDst decay can only occur via the \W-exchange or penguin-annihilation diagram.}
    \label{fig:feynman}
\end{figure}

Decays of \B mesons into two charm mesons can be used to probe elements of the \ac{CKM} matrix \cite{Cabibbo:1963yz,Kobayashi:1973fv}. Measurements of \CP violation in \mbox{\decay{\Bd}{\D^{(\ast)+}\D^{(\ast)-}}}, \mbox{\decay{\Bs}{\D^{(\ast)+}\D^{(\ast)-}}} and \mbox{\decay{\Bs}{\D_\squark^{(\ast)+}\D_\squark^{(\ast)-}}} decays can be used to determine the \ac{CKM} angles $\beta$~\cite{Aubert:2006ia,Aubert:2008ah,Fratina:2007zk,Rohrken:2012ta,LHCb-PAPER-2016-037,LHCb-PAPER-2019-036} and $\betas$~\cite{LHCb-PAPER-2014-051}, although \CP violation in \mbox{\decay{\Bs}{\D^{\ast+}\D^{\ast-}}} and \mbox{\decay{\Bs}{\D_\squark^{\ast+}\D_\squark^{\ast-}}} has not yet been measured.
These determinations are affected by higher-order Standard Model effects\cite{ALEKSAN1993173,Sanda:1996pm,Xing:1998ca,Xing:1999yx,Pham:1999fy}, which can be constrained using measurements of \CP violation parameters or branching ratios in additional decays of the \BToDD family, \eg the \BsToDstDst decay\footnote{Charge conjugation is implied throughout the paper.} \cite{Jung:2014jfa,Bel:2015wha}. In the \BsToDstDst decay, tree and penguin transitions cannot contribute. This makes these decays sensitive to the effects of \W-exchange and penguin-annihilation diagrams. The dominant Feynman diagrams of the family of \BToDD decays are shown in \cref{fig:feynman}. The absolute branching fraction of the \BsToDstDst decay is predicted to be $(3.1^{+1.3}_{-1.1})\times 10^{-4}$ using a perturbative QCD approach based on $k_T$ factorisation \cite{Li:2009xf}.

In this paper, the first observation of the \BsToDstDst decay is presented.
Its branching fraction is measured relative to that of the \BdToDstDst decay, thereby canceling systematic uncertainties originating from the uncertainty on the integrated luminosity and the \bbbar cross section. 
The measurement is performed with data collected by the \ac{LHCb} experiment, where proton beams collide at centre-of-mass energies up to $13\tev$.
In 2011 and 2012 data samples corresponding to $1\invfb$ and $2\invfb$ were collected at centre-of-mass energies of $7$ and $8\tev$, respectively, and from 2015 to 2018 a data sample corresponding to $6\invfb$ was collected at $13\tev$.
Due to the different centre-of-mass energies and due to different detector settings, the analysis is performed separately for the two data-taking periods 2011--2012 (Run~1) and 2015--2018 (Run~2), combining both results in the end. 

The measurement of the ratio of branching fractions relies on the calculation of the ratio of selection efficiencies and the determination of the signal yields for the two decays. Therefore, the analysis begins with the selection of signal and control mode candidates in \cref{sec:SelectionCandidates} and continues with the extraction of the \Bs and \Bd yields using mass fits in \cref{sec:ExtractionSignalYields}. The systematic uncertainties are evaluated in \cref{sec:StudiesSystematicUncertainties} and the final calculation of the ratio of branching fractions is presented in \cref{sec:results}.

\section{Detector and simulation}
\label{sec:Detector}
The \ac{LHCb} detector~\cite{LHCb-DP-2008-001,LHCb-DP-2014-002} is a single-arm forward
spectrometer covering the \mbox{pseudorapidity} range $2<\eta <5$,
designed for the study of particles containing \bquark or \cquark
quarks. The detector includes a high-precision tracking system
consisting of a silicon-strip vertex detector surrounding the proton-proton ($pp$)
interaction region, a large-area silicon-strip detector located
upstream of a dipole magnet with a bending power of about
$4{\mathrm{\,Tm}}$, and three stations of silicon-strip detectors and straw
drift tubes placed downstream of the magnet.
The tracking system provides a measurement of the momentum, \ptot, of charged particles with
a relative uncertainty that varies from 0.5\% at low momentum to 1.0\% at 200\gevc.
The minimum distance of a track to a \ac{PV}, the \ac{IP}, 
is measured with a resolution of $(15+29/\pt)\mum$,
where \pt is the component of the momentum transverse to the beam, in\,\gevc.
Different types of charged hadrons are distinguished using information
from two ring-imaging Cherenkov detectors. 
Photons, electrons and hadrons are identified by a calorimeter system consisting of
scintillating-pad and preshower detectors, an electromagnetic
and a hadronic calorimeter. Muons are identified by a
system composed of alternating layers of iron and multiwire
proportional chambers.

Simulation is used to calculate the selection efficiencies and to determine the shapes of the mass distributions.
In the simulation, $pp$ collisions are generated using \pythia~\cite{Sjostrand:2007gs} 
with a specific \ac{LHCb} configuration~\cite{LHCb-PROC-2010-056}.
The \evtgen package~\cite{Lange:2001uf} is used to decay unstable particles, with final-state radiation generated using \photos~\cite{davidson2015photos}.
The interaction of the generated particles with the detector, and its response,
are implemented using the \geant
toolkit~\cite{Allison:2006ve, Agostinelli:2002hh} as described in
Ref.~\cite{LHCb-PROC-2011-006}. 

\section{Selection of candidates}
\label{sec:SelectionCandidates}

The online event selection is performed by a trigger \cite{LHCb-DP-2012-004, LHCb-DP-2019-001}, which consists of a hardware stage, based on information from the calorimeter and muon systems, followed by a software stage, which applies a full event reconstruction.
At the hardware trigger stage, events are required to have a muon with high \pt or a hadron, photon or electron with high transverse energy in the calorimeters. 
The software trigger requires a two-, three- or four-track secondary vertex with a significant displacement from any primary $pp$ interaction vertex. At least one charged particle must have a large transverse momentum and be
inconsistent with originating from a \ac{PV}.
A multivariate algorithm~\cite{BBDT,LHCb-PROC-2015-018} is used for
the identification of secondary vertices consistent with the decay
of a \bquark hadron.

The \BsToDstDst and \BdToDstDst decays are reconstructed through \decay{\Dstarp}{\Dz\pip} decays, where the \Dz meson decays into the final states \Km\pip, \Kp\Km or \pip\pim.
These are the dominant two-body \Dz decays with charged final-state particles.
Several selection requirements are applied to the final-state tracks, including having a small probability of being a ghost track\cite{DeCian:2017ytk}, good track quality and loose \ac{PID} criteria.
The final-state particles and the intermediate \Dz and \Dstar mesons\footnote{Throughout this paper \Dstar and \Dz are used to refer to the \Dstarp and \Dstarm and to the \Dz and \Dzb mesons.} are required to have a high \chisqip with respect to any \ac{PV}, where \chisqip is defined as the difference in the vertex-fit \chisq of a given \ac{PV} reconstructed with and without the candidate being considered. 
The distance of closest approach between all possible combinations of tracks forming $D$ vertices has to be smaller than $0.5\mm$.
The cosine of the angle between the momentum of each $D$ meson and the direction from the best \ac{PV} to the decay vertex has to be greater than zero, where the best \ac{PV} is the \ac{PV} for which the \B candidate\footnote{Throughout this paper a \B candidate is used to refer to a possible \Bs or \Bd candidate.} has the lowest \chisqip.

The differences between the reconstructed \Dstar and \Dz invariant masses and their known values \cite{PDG2020} have to be smaller than $50\mevcc$ and $100\mevcc$, respectively.
The reconstructed mass difference between the \Dstar and \Dz candidates is required to be smaller than $150\mevcc$. Both \Dstar mesons must originate from a common vertex, and the \B candidate formed must have \pt greater than $5000\mevc$. 
The \B candidates are required to have a \chisqip smaller than 25  with respect to the best \ac{PV}.
The decay time of the \B meson has to be larger than $0.2\ps$ to suppress background from particles produced directly in the \ac{PV}. As a final requirement, a kinematic fit~\cite{Hulsbergen:2005pu} is applied to the decay chain, where the masses of the $D$ mesons and the \ac{PV} are constrained, and the resulting \chisq must be less than $50$. This fit is used to correctly account for correlations and uncertainties of the vertex positions of the \Dstar and \Dz mesons, particle momenta, flight distances, decay times, and invariant masses. 
The mass estimate from this constrained fit improves the $B$ mass resolution by around $50\%$, hence this mass estimate is also used for the extraction of signal candidates in \cref{sec:ExtractionSignalYields}.

A high fraction of events with multiple signal candidates is found in the data, even though the predicted branching fraction is too low for these to be true independent decays.
Those are classified into two categories.
Multiple candidates that originate from misidentification of a final-state particle occur in $\NumberWithMisIDRunOne\%$ and $\NumberWithMisIDRunTwo\%$ of all events for Run~1 and Run~2, respectively. 
These candidates are identified by having the same tracks for all final state particles, but where one \kaon has been misidentified as a \pion or vice versa.
For this category, the candidate corresponding to the decay with the largest branching ratio is chosen. This method is used since it is expected that most of the real decays are present in the \decay{\Dz}{\Km\pip} channel.
Multiple candidates that involve unrelated final-state particles occur in $\NumberWithoutMisIDRunOne\%$ and $\NumberWithoutMisIDRunTwo\%$ of all events for Run~1 and Run~2, respectively, and are rejected randomly so that only one candidate remains per event.
\section{Extraction of signal yields}
\label{sec:ExtractionSignalYields}

Extended maximum-likelihood fits to the distribution of the \Dstarp\Dstarm invariant mass $m_{\Dstarp\Dstarm}$ are performed to extract the yields of the \BsToDstDst and \BdToDstDst decays.
The mass fits are performed separately for data collected in Run~1 and Run~2. The fits are restricted to the mass range \mbox{$5100 < m_{\Dstarp\Dstarm} < 6000 \mevcc$}.
The choice of this mass range reduces partially-reconstructed backgrounds from \decay{\B^{0(+)}}{\Dstarp\Dstarm\pion^{0(+)}} decays, which may involve intermediate \decay{D_1(2420)^{0(+)}}{\Dstarp \pion^{-(0)}} or $\mbox{\decay{D_1(2430)^{0(+)}}{\Dstarp \pion^{-(0)}}}$ decays,  to a negligible level.
This is validated using simulated samples of these possible backgrounds generated with the RapidSim package \cite{cowan:2016tnm}.

The total \ac{PDF} consists of the signal components for the \BsToDstDst and \BdToDstDst decays and a component for the combinatorial background, \ie,
\begin{align}
 \mathcal{P} = \yield{}{\Bd} \pdf{}{\Bd} +  \yield{}{\Bs} \pdf{}{\Bs} +\yield{}{Comb} \pdf{}{Comb} \, ,
\end{align}
where the \ac{PDF} and yield of each decay mode is given by $\mathcal{P}_i$ and $N_i$, respectively. The \Bs and \Bd signal PDFs both consist of a single double-sided Hypatia function \cite{Santos:2013gra}.
The  mean value $\mu$ of the \Bs signal \ac{PDF} is shifted relative to the mean value of the \Bd \ac{PDF} by the fixed value $\mu_\Bs - \mu_{\Bd} = 87.26\mevcc$ calculated with the known meson masses \cite{PDG2020}.
Due to the limited number of selected events in data and simulation, the parameters $\zeta$ and $\beta$ of the generalised hyperbolic distribution $G(m-\mu,\,\sigma,\,\lambda,\,\zeta,\,\beta)$ of the Hypatia functions are set to zero to achieve a more stable fit.
The simulated samples only contain decays with \decay{\Dz}{\Km\pip} final states, but
the same mass model is used for all final states, as any difference between them is expected to be negligible due to the constraints on both \Dstar masses.
The combinatorial background is described with a \ac{PDF} consisting of an exponential function with a negative slope.
The same mass model is used for both data-taking periods with different sets of parameters.

\begin{figure}
  \centering
  \includegraphics[width=0.49\textwidth]{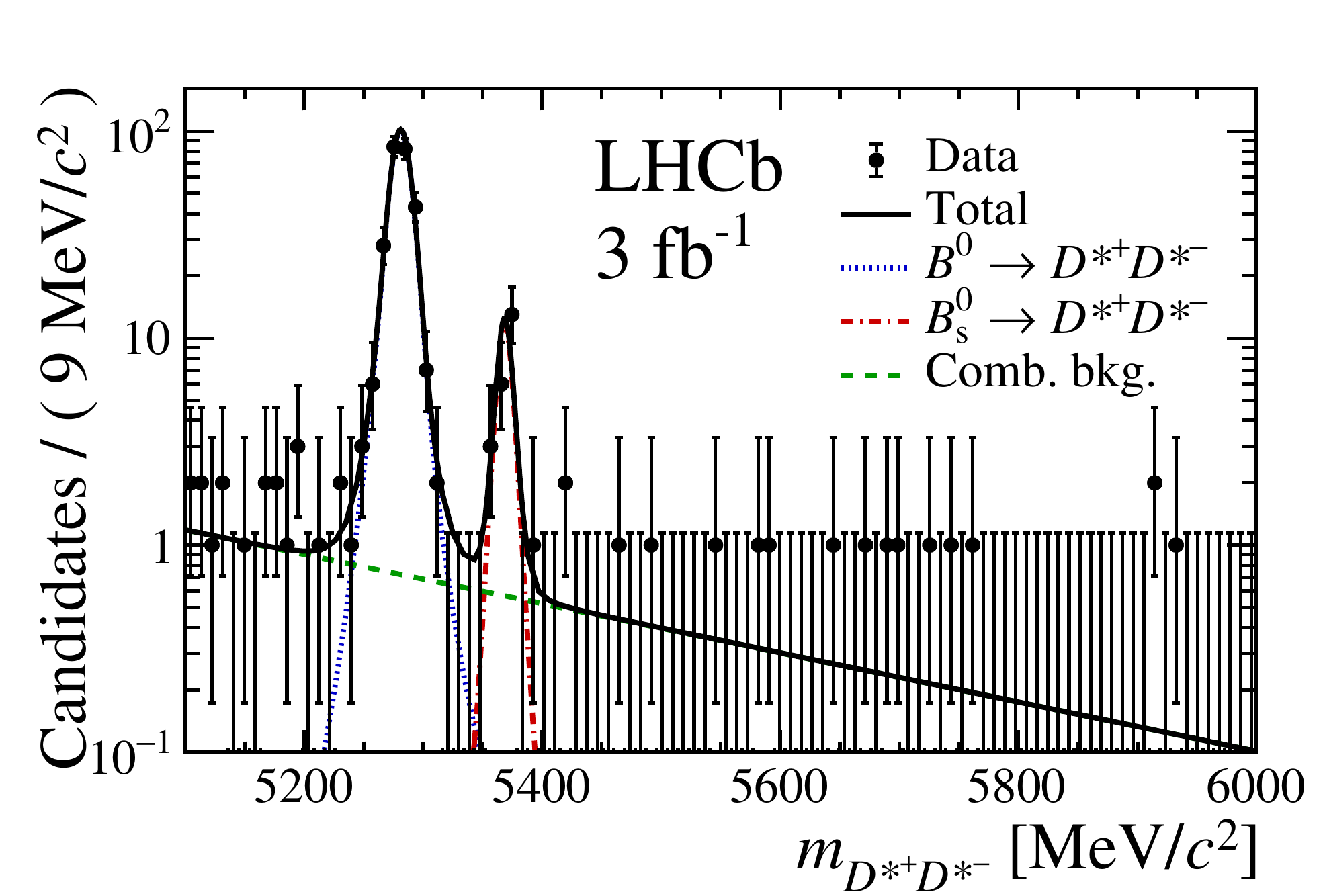}
  \includegraphics[width=0.49\textwidth]{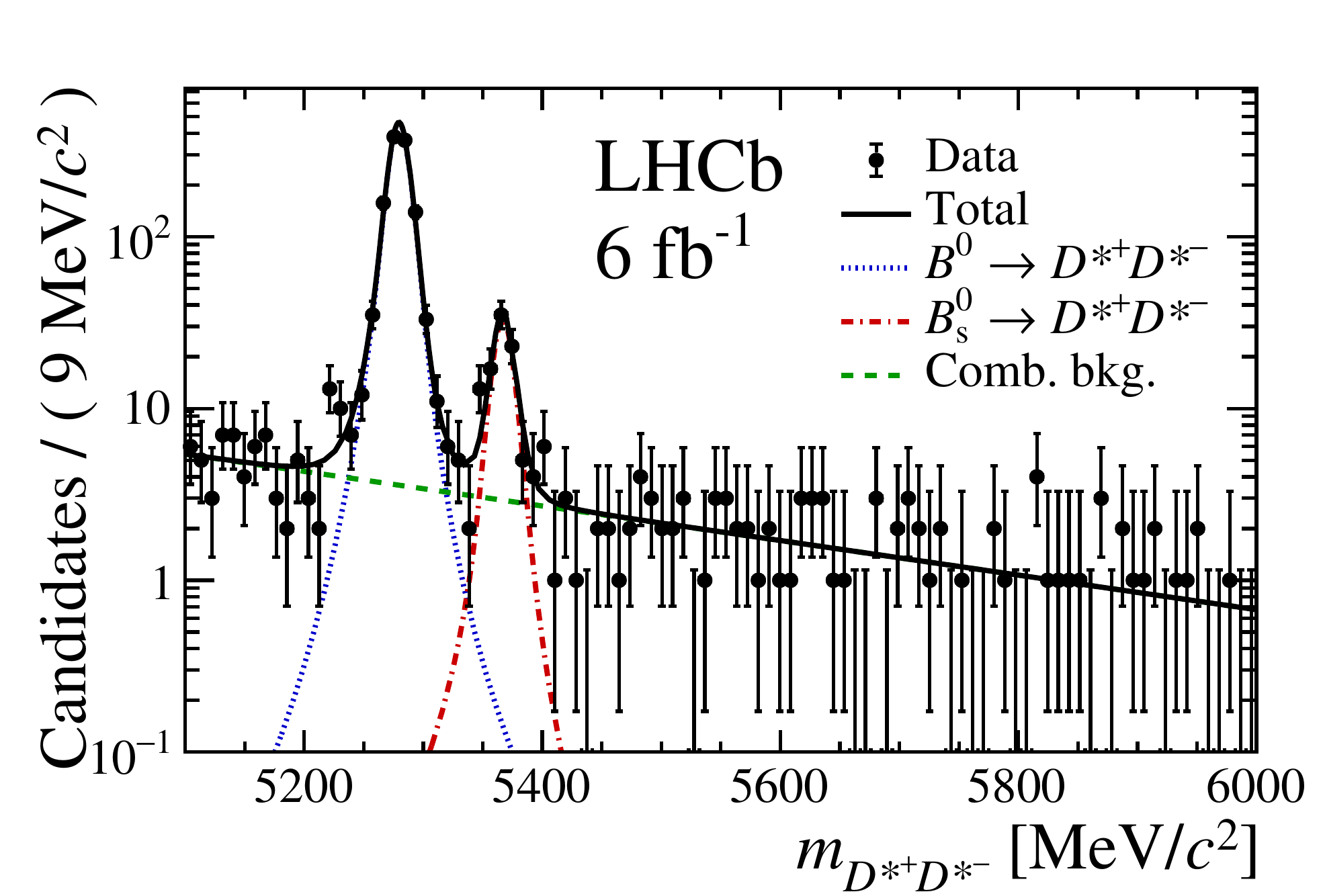}\\
  \caption{Distributions of the \Dstarp\Dstarm mass for (left) Run~1 and (right) Run~2 data samples, with the fit results overlaid.
  In addition to the full fit result (solid black), each of the individual components are shown: the \BdToDstDst signal (dotted blue), the \BsToDstDst signal (dash-dotted red) and the combinatorial background (dashed green).}
  \label{fig:mass_fit}
\end{figure}

In the fit to the data, the only free parameters are the mean of the \Bd \ac{PDF}, the widths of both signal \acp{PDF}, the slope of the exponential function and the yield of each contribution.
The mass distributions and fit projections are shown in \cref{fig:mass_fit} together with the contribution of each component.
The resulting yields for the \mbox{\BsToDstDst} and \mbox{\BdToDstDst} decays are $\BsNominalRunOne$ and $\BdNominalRunOne$, respectively, for the data-taking period Run 1. For Run 2 the \mbox{\BsToDstDst} and \mbox{\BdToDstDst} yields are $\BsNominalRunTwo$ and $\BdNominalRunTwo$, respectively. For both data-taking periods, the uncertainties are purely statistical.

\section{Systematic uncertainties}
\label{sec:StudiesSystematicUncertainties}

The systematic uncertainties are summarised in \cref{tab:systematics}.
The combined uncertainties are calculated from the individual uncertainties for Run~1 and Run~2, taking into account the correlations.
The individual systematic uncertainties are combined in quadrature to obtain the total uncertainties for each data-taking period.

\begin{table}
  \centering
  \caption{Systematic uncertainties relative to the measured value of the ratio of branching fractions for Run~1, Run~2, and their combination, taking into account correlations.}
  \label{tab:systematics}
  \begin{tabular}{lccc} 
    \toprule
    Source    &     Run~1 $[\%]$ & Run~2 $[\%]$ & Combined $[\%]$\\
    \midrule
    Size of simulated sample&1.50&1.09&0.93 \\ 
    Polarisation&4.47&3.66&3.82 \\ 
    Multiple candidates&0.12&1.05&0.85 \\ 
    Mass model&0.08&0.09&0.09 \\ 
    Mass fit range&1.49&0.51&0.70 \\ 
    \midrule
    Total without $f_\squark/f_\dquark$ & 4.94 & 3.99 & 4.08 \\
    $f_\squark/f_\dquark$ &3.18&3.11&3.11 \\ 
    \midrule
    Total & 5.88 & 5.06 & 5.13 \\
    \bottomrule
  \end{tabular}
\end{table}

The measurement of the ratio of branching fractions relies on the ratio of \bquark quark hadronisation fractions to \Bs and \Bd mesons, $f_\squark/f_\dquark$, which is taken from Ref.~\cite{LHCb-PAPER-2020-046}.
For the data-taking periods Run~1 and Run~2, all uncertainties of $f_\squark/f_\dquark$ are shared between the samples with one additional independent uncertainty of around $0.7\%$ for the Run 1 sample, so overall the uncertainties are partially correlated when the two data-taking periods are combined.

Two sources of systematic uncertainty on the efficiency ratio of the \BsToDstDst and \BdToDstDst modes are considered.
The first comes from the limited size of the simulated sample and is treated as uncorrelated between the data-taking periods.
Secondly, the nominal efficiency ratio is calculated using simulated unpolarised \BsToDstDst and \BdToDstDst decays, but these decays involve a pseudoscalar decaying into two vector mesons, which have to be longitudinally or transversely polarised.
The effect of different polarisations of the \Bd and \Bs decay products are analysed by recalculating the efficiency ratio using polarised simulated samples.
The polarisation fractions of the \Bd decay are set to their measured values \cite{Kronenbitter:2012ha}.
Different polarisation configurations for the \Bs decay products, ranging from purely longitudinal to purely transverse, are evaluated.
The largest deviation to the nominal efficiency ratio is assigned as a systematic uncertainty.
The uncertainties of the two data-taking periods are assumed to be fully correlated.

The fraction of events with multiple candidates originating from particle misidentification is of order $2\%$.
In addition to the branching-fraction based method described above, two alternative rejection methods for this kind of multiple candidates are studied to assign a systematic uncertainty:
random rejection and rejection according to a probability based on the \ac{PID} response.
For the latter, a ratio for each final state particle of the \Dz decays is calculated indicating how likely the particle is correctly identified.
The mass fit is repeated using the alternative rejection methods to extract the new yields.
The highest deviation from the default method or between the two alternative methods is assigned as a systematic uncertainty. The uncertainties of the 
two data-taking periods are assumed to be uncorrelated.

Finally, systematic uncertainties are assigned to account for the choice of the \ac{PDF} used to describe the signal distributions and to account for the choice of the mass range in the fit.
Both uncertainties are estimated using pseudoexperiments.
Data sets are generated using alternative mass models of the \Bd and \Bs contributions, or using alternative mass-fit ranges.
The alternative mass model consists of a sum of two Crystal Ball functions\cite{Skwarnicki:1986xj}, which have tails on opposite sides of the peak and parameters that are obtained using fits to simulated samples.
The remaining parameters are generated according to their results from the default mass fit.
The alternative mass ranges have the upper boundary of the default fit, but the lower boundary is shifted by $\pm50\mevcc$. 
The parameters are generated using the default mass fit results for the signal contributions and results of fits in the respective ranges for the background contribution. 
For each generated data set, an extended maximum likelihood fit is performed using the default model.
In the investigation of the alternative mass ranges, the fit uses the respective alternative ranges instead of the default fit range.
The deviation of the ratio of the yields relative to the default ratio is calculated for each data set and the mean of the deviations is assigned as the systematic uncertainty.
The results are assumed to be fully correlated between the two data-taking periods.

\section{Results}
\label{sec:results}
The \BsToDstDst decay is observed with more than $10$ standard deviations, determined using Wilks' theorem \cite{Wilks:1938dza}.

The ratio of branching fractions is given by
\begin{align}
  \frac{\mathcal{B}(\decay{\Bs}{\Dstarp\Dstarm})}{\mathcal{B}(\decay{\Bd}{\Dstarp\Dstarm})} =
  \frac{\yield{}{\Bs}}{\yield{}{\Bd}} \frac{\varepsilon_{\Bd}}{\varepsilon_{\Bs}}
  \frac{f_d}{f_s} \,,
  \label{eq:branching_ratio}
\end{align}
where $\varepsilon_{\Bd}$ and $\varepsilon_{\Bs}$ are the efficiencies of the entire selection, $\yield{}{\Bd}$ and $\yield{}{\Bs}$ are yields determined from the fit to the \Dstarp\Dstarm mass distribution, and $f_\squark/f_\dquark$ is the ratio of hadronisation fractions of \Bd and \Bs, which is taken from Ref.~\cite{LHCb-PAPER-2020-046}. 

Accordingly, the $f_\squark/f_\dquark$ ratios are given by
\begin{align*}
	\left.{\frac{f_\squark}{f_\dquark}}\right\vert_\text{Run 1} =  \fsoverfdRunOne \,,\qquad
	\left.{\frac{f_\squark}{f_\dquark}}\right\vert_\text{Run 2} =  \fsoverfdRunTwo \, .
\end{align*} 
Calculating the selection efficiencies using simulated samples results in 
\begin{align*}
	\left.{\frac{\varepsilon_\Bs}{\varepsilon_\Bd} }\right\vert_\text{Run 1}  =  \EfficiencyRatioRunOne \,,\qquad
	\left.{\frac{\varepsilon_\Bs}{\varepsilon_\Bd} }\right\vert_\text{Run 2} =  \EfficiencyRatioRunTwo \, .
\end{align*} 
The ratios of the numbers of signal candidates resulting from the fit are calculated to be
\begin{align*}
	\left.{\frac{\yield{}{\Bs}}{\yield{}{\Bd}} }\right\vert_\text{Run 1} =  \CandidatesRatioRunOne \,,\qquad
	\left.{\frac{\yield{}{\Bs}}{\yield{}{\Bd}} }\right\vert_\text{Run 2} =  \CandidatesRatioRunTwo \, .
\end{align*}  
The relative branching fractions for Run~1 and Run~2 are determined to be
\begin{align*}
  \left.{\frac{\mathcal{B}(\decay{\Bs}{\Dstarp\Dstarm})}{\mathcal{B}(\decay{\Bd}{\Dstarp\Dstarm})}}\right\vert_\text{Run 1} &= \finalbrRunOneValue \pm \finalbrRunOneStat \pm \finalbrRunOneSystNoFsFd \pm \finalbrRunOneSystOnlyFsFd \,,\\
  \left.{\frac{\mathcal{B}(\decay{\Bs}{\Dstarp\Dstarm})}{\mathcal{B}(\decay{\Bd}{\Dstarp\Dstarm})}}\right\vert_\text{Run 2} &= \finalbrRunTwoValue \pm \finalbrRunTwoStat \pm \finalbrRunTwoSystNoFsFd \pm \finalbrRunTwoSystOnlyFsFd\,,
\end{align*}
where the first uncertainty is statistical, the second is systematic and the third is due to the uncertainty of the fragmentation fraction ratio $f_\squark/f_\dquark$.
The results are combined by calculating a weighted average, using the total uncertainties on the individual measurements and taking into account correlations.
The combined ratio of branching fractions is
\begin{align*}
  \frac{\mathcal{B}(\decay{\Bs}{\Dstarp\Dstarm})}{\mathcal{B}(\decay{\Bd}{\Dstarp\Dstarm})} = \finalbrCombinedValue \pm \finalbrCombinedStat \pm \finalbrCombinedSystNoFsFd \pm \finalbrCombinedSystOnlyFsFd\, .
\end{align*}
Using the known value of the branching fraction of the \BdToDstDst decay \cite{PDG2020}, the \BsToDstDst branching fraction is calculated to be
\begin{align*}
  \mathcal{B}(\decay{\Bs}{\Dstarp\Dstarm}) = (2.15 \pm 0.26 \pm 0.09 \pm 0.06 \pm 0.16)\times 10^{-4} \,,
\end{align*}
where the fourth uncertainty is due to the \BdToDstDst branching fraction. 

The central value of the time-integrated branching fraction presented in this paper depends on the lifetime of the \Bs meson~\cite{DeBruyn:2012wj}. There is currently no measurement or prediction for the \Bs lifetime in this decay, so the average \Bs lifetime is used in the generation of simulated \BsToDstDst decays. However, the light and heavy eigenstates have significantly different lifetimes,  $\tau(B^0_{\squark,L}) = 1.423\ps$ and $\tau(B^0_{\squark,H}) = 1.620\ps$~\cite{PDG2020}. Since the \Bs selection efficiency depends on the lifetime, correction factors for the efficiency are calculated for the scenarios where either a purely heavy or a purely light \Bs eigenstate is considered. The time-integrated correction factors for the efficiency are $\integratedcL$ and $\integratedcH$ for light and heavy \Bs eigenstates, respectively. If the lifetime of the \Bs meson in the \BsToDstDst decay is found in further studies to differ significantly from the average \Bs lifetime, the central value of the paper result can be corrected. The equivalent effect also occurs for the \BdToDstDst decay but is neglected due to the small value of $\Updelta\Gamma_\dquark$\cite{PDG2020}.

\section{Summary}
\label{sec:Summary}

The \BsToDstDst decay is observed with more than $10$ standard deviations.
Its branching fraction is measured relative to the \BdToDstDst branching fraction.
The measurement is performed using data collected by the \ac{LHCb} experiment in Run~1 and Run~2 corresponding to an integrated luminosity of $9\invfb$. 
The time-integrated ratio of branching fractions is measured to be
\begin{align*}
  \frac{\mathcal{B}(\decay{\Bs}{\Dstarp\Dstarm})}{\mathcal{B}(\decay{\Bd}{\Dstarp\Dstarm})} = \finalbrCombinedValue \pm \finalbrCombinedStat \pm \finalbrCombinedSystNoFsFd \pm \finalbrCombinedSystOnlyFsFd\, ,
\end{align*}
where the first uncertainty is statistical, the second systematic and the third is due to the uncertainty of the fragmentation fraction ratio $f_\squark/f_\dquark$. The ratio is measured assuming an average \Bs lifetime in the calculation of the \Bs selection efficiency, since no measurement or prediction for the \Bs lifetime is available in this decay. Correction factors for the efficiency are reported in \cref{sec:results} for the scenarios where either a purely heavy or a purely light \Bs eigenstate is considered, so that the central value of the result presented here can be recalculated when a measurement or prediction is available.
The \BsToDstDst branching fraction is determined to be
\begin{align*}
  \mathcal{B}(\decay{\Bs}{\Dstarp\Dstarm}) = (2.15 \pm 0.26 \pm 0.09 \pm 0.06 \pm 0.16)\times 10^{-4} \,,
\end{align*}
where the fourth uncertainty is due to the \BdToDstDst branching fraction.
The result is in agreement with the theoretical prediction using a perturbative QCD approach based on $k_T$ factorisation\cite{Li:2009xf}.
It can be used to constrain higher-order SM effects in \CP violation observables in \decay{\Bd}{\D^{(\ast)+}\D^{(\ast)-}} and \decay{\Bs}{\D_\squark^{(\ast)+}\D_\squark^{(\ast)-}} decays.

\section*{Acknowledgements}
%
%
\noindent We express our gratitude to our colleagues in the CERN
accelerator departments for the excellent performance of the LHC. We
thank the technical and administrative staff at the LHCb
institutes.
We acknowledge support from CERN and from the national agencies:
CAPES, CNPq, FAPERJ and FINEP (Brazil); 
MOST and NSFC (China); 
CNRS/IN2P3 (France); 
BMBF, DFG and MPG (Germany); 
INFN (Italy); 
NWO (Netherlands); 
MNiSW and NCN (Poland); 
MEN/IFA (Romania); 
MICINN (Spain); 
SNSF and SER (Switzerland); 
NASU (Ukraine); 
STFC (United Kingdom); 
DOE NP and NSF (USA).
We acknowledge the computing resources that are provided by CERN, IN2P3
(France), KIT and DESY (Germany), INFN (Italy), SURF (Netherlands),
PIC (Spain), GridPP (United Kingdom), 
CSCS (Switzerland), IFIN-HH (Romania), CBPF (Brazil),
Polish WLCG  (Poland) and NERSC (USA).
We are indebted to the communities behind the multiple open-source
software packages on which we depend.
Individual groups or members have received support from
ARC and ARDC (Australia);
Minciencias (Colombia);
AvH Foundation (Germany);
EPLANET, Marie Sk\l{}odowska-Curie Actions and ERC (European Union);
A*MIDEX, ANR, IPhU and Labex P2IO, and R\'{e}gion Auvergne-Rh\^{o}ne-Alpes (France);
Key Research Program of Frontier Sciences of CAS, CAS PIFI, CAS CCEPP, 
Fundamental Research Funds for the Central Universities, 
and Sci. \& Tech. Program of Guangzhou (China);
GVA, XuntaGal, GENCAT and Prog.~Atracci\'on Talento, CM (Spain);
SRC (Sweden);
the Leverhulme Trust, the Royal Society
 and UKRI (United Kingdom).

\addcontentsline{toc}{section}{References}
\bibliographystyle{LHCb}
\bibliography{main,standard,LHCb-PAPER,LHCb-CONF,LHCb-DP,LHCb-TDR}

\newpage
\centerline
{\large\bf LHCb collaboration}
\begin
{flushleft}
\small
R.~Aaij$^{32}$\lhcborcid{0000-0003-0533-1952},
A.S.W.~Abdelmotteleb$^{50}$\lhcborcid{0000-0001-7905-0542},
C.~Abellan~Beteta$^{44}$,
F.~Abudin{\'e}n$^{50}$\lhcborcid{0000-0002-6737-3528},
T.~Ackernley$^{54}$\lhcborcid{0000-0002-5951-3498},
B.~Adeva$^{40}$\lhcborcid{0000-0001-9756-3712},
M.~Adinolfi$^{48}$\lhcborcid{0000-0002-1326-1264},
P.~Adlarson$^{77}$\lhcborcid{0000-0001-6280-3851},
H.~Afsharnia$^{9}$,
C.~Agapopoulou$^{13}$\lhcborcid{0000-0002-2368-0147}
% ...

{\footnotesize \it

% ...
$^{9}$Universit{\'e} Clermont Auvergne, CNRS/IN2P3, LPC, Clermont-Ferrand, France\\
% ...
$^{13}$LPNHE, Sorbonne Universit{\'e}, Paris Diderot Sorbonne Paris Cit{\'e}, CNRS/IN2P3, Paris, France\\
% ...
$^{32}$Nikhef National Institute for Subatomic Physics, Amsterdam, Netherlands\\
% ...
$^{40}$Instituto Galego de F{\'\i}sica de Altas Enerx{\'\i}as (IGFAE), Universidade de Santiago de Compostela, Santiago de Compostela, Spain\\
% ...
$^{44}$Physik-Institut, Universit{\"a}t Z{\"u}rich, Z{\"u}rich, Switzerland\\
% ...
$^{48}$H.H. Wills Physics Laboratory, University of Bristol, Bristol, United Kingdom\\
% ...
$^{50}$Department of Physics, University of Warwick, Coventry, United Kingdom\\
% ...
$^{53}$School of Physics and Astronomy, University of Glasgow, Glasgow, United Kingdom\\
$^{54}$Oliver Lodge Laboratory, University of Liverpool, Liverpool, United Kingdom\\
% ...
$^{77}$Department of Physics and Astronomy, Uppsala University, Uppsala, Sweden, associated to $^{53}$\\
% ...
}
\end{flushleft}

\end{document}